\documentstyle[12pt]{article}

\begin{document}

\centerline{\Large On the Indication from Pioneer 10/11 Data of an Anomalous Acceleration}

\bigskip\bigskip

\centerline{Yong Gwan Yi}

\bigskip

\begin{abstract}
Hubble's law, which states a linear increase in velocities with
distances, can physically be understood in terms of an acceleration
$cH$. This work proposes a connection between this ``universal''
acceleration seen in the solar system and the anomalous acceleration
acting on the Pioneer 10/11 spacecraft, in which the Hubble constant
inferred from Pioneer 10/11 data is $\sim 87$ km/s/Mpc. Its physical
implication is discussed in relation with Mach's principle.
\end{abstract}

\bigskip\bigskip

By 1998, when Pioneer 10 was 71 AU away from the Sun, one team of
researchers at the tracking stations published that radio metric data from
Pioneer 10/11 had indicated an apparent anomalous acceleration acting
on the spacecraft with a magnitude $\sim 8.5\times 10^{-8}$ cm/s$^2$,
directed towards the Sun [1]. When Pioneer 10 ventured beyond
the realm of the planetary system, Anderson {\it et al}. began
monitoring its orbit for evidence of the long-hypothesized Planet X.
They found no such planet, but they did notice some extra tiny slowing
of its outward motion. Beginning in 1980, when at 20 AU the solar
radiation pressure acceleration had decreased to $<5\times 10^{-8}$
cm/s$^2$, Jet Propulsion Laboratory's orbit determination program
analysis of Pioneer 10/11 data found the biggest systematic error
in the acceleration residuals. Even after all known sources of gravity
and other forces were taken into account, the apparent acceleration
seemed to be present in the residuals.

Ever since the effect was reported, there has been intense
debate over its origin. Murphy [2] proposes that the anomalous
acceleration can be explained, at least in part, by nonisotropic
radiative cooling of the spacecraft electronics. Katz [3] argues
that the anomalous acceleration may be due to anisotropic
heat reflection off of the back of the antenna dish.
Anderson {\it et al}. [4] respond, these explanations fall short
of accounting for the anomalous Pioneer 10 acceleration. But
a few of them suppose a gas leak from thruster to be its origin.
Scheffer [5] asserts that the proposed mechanism much
more likely explains the anomaly. Meanwhile, it is noted that
the size of the anomaly is of the order of $cH$, where $H$
is the Hubble constant. Rosales [6] and \"{O}stvang [7]
make attempts to develop a space-time metric which
incorporates the effect of cosmic expansion. Nottale [8]
tries to tie this to the cosmological constant at the scale of
the solar system. Independent of the note, I have come to
see an acceleration $cH$ in connection with the Pioneer effect.
I should like to show a possible account of the anomalous
acceleration on physical considerations.

In attempts to explain the effect, my attention focused on the
fact that the solar system rotating with the Galactic rotation has
a centrifugal acceleration of $\sim 1.8 \times 10^{-8}$ cm/s$^2$,
the same order of magnitude. Moreover, the centrifugal acceleration
was consistent with observation that no magnitude variation of the
acceleration with distance was found, within a sensitivity of $2
\times 10^{-8}$ cm/s$^2$ over a range of 40 to 60 AU. The points
led me to put the weight of its possible explanation in the motion
of the solar system.

Non-uniform rotation of our Galaxy gives a hint on its internal motions
such as local expansion or contraction while rotating, making an
additional contribution to the centrifugal acceleration. It can be
estimated using the experimental curve of the rotating velocity versus
the distance from the axis [9]. In the curve the gradient of velocity
at the position of the solar system is seen to be about $-10$ km/s/kpc,
by which non-uniform rotation makes one order of magnitude small
contribution to the centrifugal acceleration [10]. The Coriolis effect
on the moving Pioneer at 12.5 km/s is about 11\% in magnitude
of the centrifugal acceleration.

As no further explanation could be deduced from the Galactic rotation,
I turned my attention to the motion of our Galaxy as a whole.
Continuing my search for acceleration, I considered with reluctance
the possibility of an acceleration in a general recession of distant 
galaxies.
It came out clearly, how the recessional velocities could have been
understood in terms of an acceleration.

The announcement by E. Hubble in 1929 of a ``roughly linear relation
between velocities and distances'' established in most astronomers'
minds a sort of bird's-eye view of a general recession of distant
galaxies. But there is a physics to be found in the linear relation.
Our information about the frequency shifts comes to us through
the observation of light emitted by distant sources. The velocity of
a source at distance $r$ is a result of velocity difference between
the source at an earlier or retarded time $t-r/c$ and the observation
point at time $t$. Physically, Hubble's relation states a roughly
linear increase in relative velocity change due to the time of
propagation $\triangle t=r/c$: $v=cH\triangle t$. It becomes
evident that the linear increase in recessional velocities with
distances is a result of longer light travel times from further
distant galaxies. Hubble's law finds a natural explanation
in terms of an acceleration $cH$.

The times of propagation permit only the evaluation of galaxies
in terms of the retarded positions and velocities. As we look further
and further out into space, we see galaxies that are presumably
younger and younger, the furthest naturally being those in the
remotest past. The linear increase in recessional velocities with
distances can therefore be put in the form of a linear decrease
in relative velocities with times up to the time of observation. The
relation between velocities and times up to the time of observation
manifests the direction of acceleration against the recession.
The general recession in deep space of distant galaxies must
be slowing down at a uniform rate.

It would be of gravitational character occurring on
a scale of the universe that the general recession of distant galaxies has been
decelerating. From this perspective the value $cH$ is identified with
the gravitational field of the universe as observed in the solar system.
This assumption seems tenable, seeing that the spherically symmetric
distribution of matter produces a constant acceleration inside the
distribution. But when we identify $cH$ as the gravitational field of
the universe, we conceive the ultimate interpretation of Slipher's
red shifts as a ``universal'' gravitational effect. This is because
the red shifts can then be understood in terms of a ``universal''
gravitational potential $cHr$. In fact, the red shift effect is an
effect of only the relative distances between sources and
observation point. From the redshift-distance relation one
can only infer that distant galaxies are in free fall; their states
of motion remain unaccounted for. In principle, there is no
objection to identifying the red shifts ultimately as gravitational
red shifts caused by the gravitational field of the universe.
In appreciating cosmological relevance of red shifts a
change in the orientation of our thought is desirable.

On the basis of the argument we see that there is a ``universal''
acceleration towards the Sun of $cH$. We must adopt an active
view---A general recession of distant galaxies is the Sun-based
astronomical observation. The solar system would respond to
the external gravitational field with the same magnitude, directed
away from the Sun. From the general recession of distant galaxies,
that is, we can realize an acceleration existing in the relative
recession of our own. Pioneer 10/11 moving away from the solar
system at the approximately constant velocity make themselves
ideal instruments to probe for an additional acceleration existing
in the solar system. To the spacecraft the equation of motion would
appear as if they are moving under the influence of its inertial force.
The anomalous acceleration that has appeared in Pioneer 10/11
tracking would be an inertial reaction to the solar system accelerated
relative to distant galaxies. In magnitude and direction their assessment
is in substantial agreement with what we should expect from Hubble's
law. Considerations lead to the conclusion that the apparent
acceleration acting on the spacecraft is a reflection of the ``universal''
acceleration as seen in the solar system, in which the Hubble
constant inferred from Pioneer 10/11 data is $\sim 87$ km/s/Mpc.

Of great interest is that the acceleration $cH$ has
already been assumed in a new law of motion devised by Milgrom
[11]. He has imputed the mass discrepancy, observed in galactic
systems, not to the presence of dark matter, but to a departure
from Newtonian dynamics below the scale of acceleration.
A success of the modified dynamics in explaining astronomical
data may be interpreted as implying a need to change the law of
inertia in the limit of small accelerations. In the previous
consideration we have identified the acceleration ultimately as
the gravitational field of the universe seen in the solar system.
The consideration of the anomalous acceleration naturally leads
to speculation about the inertial reference frame defined by the
solar system. The issue of inertia piques curiosity.

One may inquire about the modification the anomalous acceleration
would assume in the solar system of Newtonian dynamics. Apparently
we are guided by a modified dynamics that imputes $cH$ to a
departure from Newtonian dynamics:
\begin{equation}
\frac{GM_{\odot}}{r^2}
\longrightarrow\frac{GM_{\odot}}{r^2}+cH.
\end{equation}
It represents an attempt to render justice to the fact that Pioneer
10/11 have been slowing down faster than predicted by Newtonian
dynamics. The modification makes it obvious that inertia is due not
only to the solar gravitational field but also to the gravitational
field of the universe. Evidently it indicates that inertial forces do
not exactly cancel solar gravitational forces for freely falling
planetary systems. The paradigm is obvious. Mach's principle
happens to be true!

Mach's principle has been the subject of some lively discussion
regarding anisotropy of inertia. Cocconi and Salpeter [12] pointed
out that there is a large mass near us, the Milky Way Galaxy, and
that Mach's principle would suggest slight differences in inertial mass
when a particle is accelerated toward or away from the Galactic center.
In the experiments [13] it was shown that with a precision of 1 part
in $10^{20}$ there is no anisotropy of inertia associated with effects
of mass in our Galaxy. Dicke [14] came to defence, arguing that as
Mach's principle associates the inertial reaction with the matter
distribution in the universe, an anisotropy in the inertial mass should
be universal, the same for all particles. I should like to add defence:
The gravitational field of the universe as observed in the solar system
is the sum of the gravitational field acting on the Milky Way and
the centrifugal acceleration due to rotation about the Milky Way,
in which the gravitational field dominates strangely somewhat.
Phenomenologically, the gravitational field of the universe as seen
in the solar system directs toward the Sun. Thus, an anisotropy of
inertia should be expected toward the Sun, and at present I am discussing such possibility from the anomalous acceleration
seen in the Pioneer 10/11 spacecraft.

Let us consider the motion
of a small body in an orbit around the Sun. The modification (1) is a phenomenological scheme which modifies
the solar system into the Newtonian frame of reference which is
compatible with Mach's principle. The added inertia to
the solar gravitational field leads to a differential equation for the
orbit of the form
\begin{equation}
\frac{d^2u}{d\theta^2}+u=
\frac{mk}{l^2}\biggl(1+\frac{mcH}{ku^2}\biggr),
\end{equation}
where $m$ is the mass of the small body, $l$ is the angular
momentum, and $u$ and $k$ denote $1/r$ and $GM_{\odot}m$.
The second term in the round bracket is the one which 
distinguishes
the solar system from the inertial frame of reference. 
Mach's principle can be formulated in this way if the anomalous 
acceleration is assumed to be the inertial reaction to the 
matter distribution in the universe.

We may solve the inertial system equation approximately.
We expand the periodic solution of the equation into a series
\begin{equation}
u=\alpha+\lambda\beta_1+\alpha \epsilon\cos(\rho\theta)
+ \lambda\sum_{n=2}^{\infty}\beta_n\cos(n\rho\theta),
\end{equation}
where $\alpha=mk/l^2$, $\lambda=mcH/k$, and $\epsilon$ is
the eccentricity of the ellipse [15]. We substitute the series
solution into the equation. For $\lambda/u^2$, we expand
\begin{eqnarray}
\frac{\lambda}{u^2}
&\sim& \frac{\lambda}{\alpha^2(1+\epsilon\cos(\rho\theta))^2}
\nonumber\\
&\sim& \frac{\lambda}{\alpha^2}\biggl(1-
2\epsilon\cos(\rho\theta)+3\epsilon^2\cos^2(\rho\theta)-
4\epsilon^3\cos^3(\rho\theta)+\cdots\quad\biggr).
\end{eqnarray}
By comparing the $\cos(\rho\theta)$ terms we obtain the equation
which determines $\rho$ to a first approximation. According to this
calculation, the elliptical orbit of a planet referred to the Newtonian
frame of reference rotates in the opposite direction as the planet
moves, with a rate that is given by
\begin{equation}
\frac{2\pi cHa^2(1-\epsilon^2)^2}{GM_{\odot}}\biggl(1+
\frac 32\epsilon^2+\frac{15}{8}\epsilon^4+\cdots\quad\biggr),
\end{equation}
where $a$ is the planetary semimajor axis.

Equation (5) describes the rate of precession at which the perihelion will have
retarded per revolution. The precession expected from Mach's principle
increases rapidly as we move away from the Sun. For Mercury it gives
the value of $10''$ per century and for Earth the value of $16.34''$.
They destroy the current agreement between the general theory of
relativity and the observed anomalous precessions. Strongly it casts
doubt on the validity of calculation. Is my calculation erroneous?
Or is there some unrecognized effect in observations?

We need to look back at the Pioneer effect. The effect could only
be seen beyond 20 AU. The anomalous acceleration acting on
Pioneer 10/11 could not be found until the solar radiation pressure
had decreased to less than a critical value. The solar
radiation pressure decreases as $r^{-2}$. As indicated for the
Pioneers, at distances $>10-15$ AU it produces an acceleration that
is much less than $8\times 10^{-8}$ cm/s$^2$, directed away from
the Sun. Hence, even granting that the rate (5) is in principle
expected, we should be aware that the inertial effect may possibly
be contributing to the motion of distant planets such as Uranus,
Neptune, and Pluto. On the motion of near planets would the inertial
effect be entirely masked by the solar radiation pressure, and
there is no prospect of its being measured.

Brans and Dicke [16] have attempted to incorporate Mach's
principle into general relativity. They suggest field equations
with a long-range scalar field produced by the total mass in the
visible universe. In line with the interpretation of Mach's principle,
the long-range scalar field matches the ``universal'' acceleration
$cH$ seen in the solar system. The modification (1) replaces the
Schwarzschild solution by its generalization
\begin{equation}
1-\frac{2}{c^2}\biggl(\frac{GM_{\odot}}{r}\biggr)
\longrightarrow
1-\frac{2}{c^2}\biggl(\frac{GM_{\odot}}{r}-cHr\biggr).
\end{equation}
We are thus led to an alternative approach by assuming that
Einstein's field equations still apply, but that the metric differs
from the Schwarzschild solution by the gravitational field of the
universe seen in the solar system. Just like an approximate
expression $gh$ for gravitational potential at height $h$ on the
Earth's surface, so will be an expression $cHr$ for gravitational
effects having their origin in the universe surrounding the solar
system. The generalization (6) introduces a new term
$-H/c$ in addition to the relativistic term in the right hand side
of (2). But it is extremely small compared to the other terms.
In their theory, Brans and Dicke make mention of the gravitational
red shift and the deflection of light in the context of Mach's principle.
In my view, however, these phenomena seem to be of optic nature
in relation to property of the medium of propagation [17].

\bigskip\bigskip

\noindent\mbox{ [1]} J D Anderson, P A Laing, E L Lau, A S Liu, 
M M Nieto, S G Turyshev,\newline\indent\mbox{ }{\it Phys. Rev. 
Lett.} {\bf 81} (1998) 2858.\newline
\mbox{ [2]} E M Murphy, {\it Phys. Rev. Lett.} {\bf 83}
(1999) 1890.\newline
\mbox{ [3]} J I Katz, {\it Phys. Rev. Lett.} {\bf 83}
(1999) 1892.\newline 
\mbox{ [4]} J D Anderson, P A Laing, E L Lau, A S Liu,
M M Nieto, S G Turyshev,\newline\indent\mbox{ }{\it Phys. Rev. 
Lett.} {\bf 83} (1999) 1891; 1893; {\it Phys. Rev.} {\bf D65}
(2002) 082004.\newline
\mbox{ [5]} L K Scheffer, {\it Phys. Rev.} {\bf D67}
(2003) 084021.\newline
\mbox{ [6]} J L Rosales, {\it e-print} gr-qc/0212019.\newline
\mbox{ [7]} D \"{O}stvang, {\it Class. Quantum Grav.}
{\bf 19} (2002) 4131.\newline
\mbox{ [8]} L Nottale, {\it e-print} gr-qc/0307042.\newline
\mbox{ [9]} D P Clemens, {\it Astrophys. J.} {\bf 295}
(1985) 422.\newline
\mbox{[10]} H Lamb, {\it Hydrodynamics} (Dover, New Yok, 1945), 
6th ed. p.28.\newline
\mbox{[11]} M Milgrom, {\it Astrophys. J.} {\bf 270} 
(1983) 365; 371; 384;\newline\indent\mbox{ }J Bekenstein, 
M Milgrom, {\it Astrophys. J.} {\bf 286} (1984) 7.\newline
\mbox{[12]} G Cocconi, E E Salpeter, {\it Nuovo Cimento} 
{\bf 10} (1958) 3608;\newline\indent\mbox{ }{\it Phys. Rev. 
Lett.} {\bf 4} (1960) 176.\newline
\mbox{[13]} V W Hughes, H G Robinson, V Beltran-Lopez, {\it 
Phys. Rev. Lett.} {\bf 4} (1960) 342;\newline\indent\mbox{ }R W P 
Drever, {\it Phil. Mag.} {\bf 6} (1961) 683.\newline
\mbox{[14]} R H Dicke, {\it Phys. Rev. Lett.} {\bf 7}
(1961) 359.\newline
\mbox{[15]} P G Bergmann, {\it Introduction to the Theory of 
Relativity}\newline\indent\mbox{ }(Prentice-Hall, New Delhi, 
1977), p.215.\newline
\mbox{[16]} C Brans, R H Dicke, {\it Phys. Rev.}
{\bf 124} (1961) 925.\newline
\mbox{[17]} Y G Yi, {\it e-print} physics/0006006.

\end{document}